**Day of the Week Effect in biotechnology stocks: An Application of the GARCH processes**

Swarn Chatterjee,* Ph.D.
205, Dawson Hall
University of Georgia, Athens, GA 30602
Email: swarn@uga.edu
Phone: 706-542-4722
**\*(Corresponding Author)**

Amy Hubble, MBA, CFA, CFP®
205, Dawson Hall
University of Georgia, Athens, GA 30602
Email: ahubble@uga.edu
Phone: 706-542-4722

Abstract

This study examines the presence of the day-of-the-week effect on daily returns of biotechnology stocks over a 16-year period from January 2002 to December 2015. Using daily returns from the NASDAQ Biotechnology Index (NBI), we find that the stock returns were the lowest on Mondays, and compared to the Mondays the stock returns were significantly higher on Wednesdays, Thursdays, and Fridays. The day-of-the-week effect on returns of biotechnology stocks remained significant even after controlling for the Fama-French and Carhart factors. Moreover, the results from using the asymmetric GARCH processes reveal that momentum and small-firm effect were positively associated with the market risk-adjusted returns of the biotechnology stocks during this period. The findings of our study suggest that active portfolio managers need to consider the day of the week, momentum, and small-firm effect when making trading decisions for biotechnology stocks. Implications for portfolio managers, small investors, scholars, and policymakers are included.

## 1. Introduction

There have been a number of studies over the past four decades that have examined the association between variations in stock market returns and the day of the trading week in the U.S. and international financial markets (Cross, 1973; French, 1980; Gibbons & Hess, 1981; Aggarwal & Rivoli, 1989; Solnik & Bousquet, 1990; Poshakwale, 1996; Doubois & Louvet, 1996; Berument & Kiymaz, 2001). Additionally, the day-of-the-week effect on asset price volatility have been globally observed for a number of different asset classes such as treasury bills and bond markets, futures markets, crude oil, and real estate (Dyl & Maberly, 1986; Bampinas, Fountas, & Panagiotidis, 2015).

Over the past two decades the biotechnology industry has been one of the fastest growing sub-sectors in the area of healthcare. These biotechnology companies focus on cutting-edge research on finding cures and remedies for complex diseases such as cancer, multiple sclerosis, rheumatoid arthritis, and various autoimmune conditions (Gitter, 2013). The products launched by these companies subsequent to their approval by the Food and Drug Administration (FDA) are amongst the most expensive and have accounted for nearly a fifth of the increase in prescription costs over the past decade (Evans, 2010). While publicly traded biotechnology companies have benefitted from tremendous innovation and research, these companies' prices in the stock markets have experienced substantial volatility and growth. Although it is possible to theoretically hypothesize the association between risk, expected return, and innovative activities of the biotechnology companies relative to the market, very little empirical literature is currently available that explains the risk and daily price movements of the stocks in the biotechnology sector. The Affordable Care Act (ACA), which was signed into law on March 23, 2010, may have long-term policy risks for the innovative biotechnology firms; although, a little more than half a decade may not be sufficient time to fully understand the effect of ACA on the biotechnology companies (Pande, 2014).

This study extends the literature by examining the day-of-the-week effect on daily returns of the NASDAQ Biotechnology Index (NBI). Implications of this study's findings for scholars of finance, traders, and individuals investors are included. The remainder of the manuscript includes a review of literature, followed by a detailed description of the methodology; the results are presented in the next section, which is followed by a discussion of the key findings of this study and its concluding implications.

## 2. Literature Review

In one of the earlies studies on the day-of-the-week effect on stock returns Cross (1973) found that between 1953 and 1970, the mean return on the Fridays was higher than the mean returns on the Mondays of the weeks during this period. French (1980) studied the day-of-the-week effect on stocks during the 1953 and 1977 period and found similar results. Similarly, Gibbons and Hess (1981) found that the mean returns on Mondays were lower than the expected returns of the Dow Jones Industrial Average stocks. Lakonishok and Levi (1982) proposed that the expected return for Mondays should be lower than that of other days as an explanation for this anomaly in day-of-the-week differences in stock returns. In other studies on international markets, Solnik and Bousquet (1990) found that the lowest returns in the Paris market were on Tuesdays; similarly Jaffe and Westerfield (1985) found that Tuesday had the lowest returns in the Australian and Japanese markets as well.

Berument and Kiymaz (2001) found that volatility in the U.S. markets was the highest on Fridays and lowest on Wednesdays. Similarly, Rodriguez (2012) found that volatility in South American markets peaked on Fridays and was the lowest on Mondays. Brusa, Liu, & Schulman (2000) found that when compared to Mondays when markets experienced a negative return, the stock returns of small firms was negative on Fridays while the stock returns of large firms was positive on the Fridays.

Chok and Sun (2007) find that the volatility of the price movements of biotechnology firms are susceptible to stock option exercise by company insiders, the age of the board members, and the extent of resource dependence of the firms (Chok & Sun, 2007). Although healthcare stocks, including the biotechnology stocks, have reacted positively to court rulings that have supported the ACA during the legislation's recent legal challenges (Kaufman, 2015), many biotechnology market experts expected the biotechnology firm profits to fall following the passage of the ACA (Douglas, 2014).

This day-of-the-week effect is inconsistent with the assumptions of efficient market hypothesis (Fama, 1965; 1970). One of the key assumptions of the efficient market hypothesis is that information is fully accounted for in the price of assets. Therefore, it would be impossible to consistently beat the market on a risk-adjusted basis. This argument is also central to the Modern Portfolio Theory and the Capital Asset Pricing Model (Fabozzi, Gupta, and Markowitz, 2002; Sharpe, 1964; Markowitz, 1952; Vernon, Golec, & DiMassi, 2009), which argue that the expected return of an asset can be predicted after controlling for U.S. treasury bills, market risk premium, the systematic risk (beta) of the stocks, and other factors suggested by Fama and French (2004).

It is, therefore, likely that the expected return for the biotechnology stocks would be same each day of the week after controlling for the risks, firm size, price movements of the market, and the general momentum of stocks. In addition, the biotechnology industry was exposed to three unique macroeconomic events that coincided with the period from March 30, 2010, through December 31, 2015. Along with the rest of the market during this period, the biotechnology industry benefited from the phase of market recovery from the great recession of 2008. This period also saw low interest rates and the Federal Reserve's quantitative easing (QE) program. But the third macroeconomic event was unique to the healthcare sector, which includes the biotechnology industry, and was as a result of the Affordable Care Act legislation of 2010. The ACA was signed into legislation on March 30, 2010. Although it is too early to fully understand the economic consequence of ACA on biotechnology stock returns, a combination of these three macroeconomic changes could possibly alter the performance of biotechnology stocks relative to the market in the period after March 30, 2010.

## 3. Methodology

### 3.1 Data

The daily information for the S&P 500 index, risk-free rate, Fama-French, and momentum factors were obtained from the Center for Research in Security Prices (CRSP) database. This study also uses the NBI. The daily price information for the NBI was obtained from the Quandl (quandl.com) database. There are 117 companies listed in the NBI. Nine of these are large company stocks, another 23 are mid-sized company stocks, and the remaining 85 are small company stocks. However, since NBI is a value-weighted index, approximately 60% of the NBI's value is comprised of the 10 large company stocks. Approximately 22% of the NBI's value is composed of the 85 small company stocks, and the remaining 18% of its total index value is comprised of medium-size company stocks. This study uses data extracted over a 14-year period starting January 1, 2002 through December 31, 2015. Additional analysis is run for the period after March 30, 2010.

### 3.2 Analyses

This paper first presents the descriptive statistics of the sample and the graphs for the price movements for the NBI between January 1, 2002 and December 31, 2015. Binary variables are constructed for the days Monday through Friday. In the empirical model, Tuesday through Friday is compared against the reference group of the Monday variable. In equation 1, the $R_{NBI}$ variable is examined for the day-of-the-week effect after risk adjustment by controlling for the additional 3 Fama and French factors (2004), and the Carhart (1997) momentum factor.

$$R_{NBI} = \beta_0 + \beta_1(Tuesday) + \beta_2(Wednesday) + \beta_3(Thursday) + \beta_4(Friday) + \sum_{l=i}^{n} \beta_5$$
$$R_{NBI-i} + f\,R_f + \beta_6(R_m - R_f) + sSMB + hHML + uUMD + e \qquad (1)$$

$R_f$ is the risk-free rate. $R_m$-$R_f$ represents the market risk premium. SMB, or 'small minus big', is the difference in the returns between the small-cap and the large-cap stocks. HML represents the difference in the returns between value stocks (High book-to-market ratio) and growth stocks (low book-to-market ratio). UMD represents the difference between the returns of stocks moving up and down in value during the period. Additional analyses with the variables are also run for the Pre-ACA (January 1, 2002 – March 30, 2010), and post-ACA (March 31, 2010 – December 31, 2015) periods.

However, when an OLS regression is run using equation 1, it assumes a constant variance of the error term. Therefore, the OLS model assumes that the error term has a '0' mean and is normally distributed. However, studies by Engle (1982) and Bollerslev (1987) found that the error term may not be constant in time-series models and could possibly vary across time. As a solution, Engle (1982) suggested the use of the Autoregressive Conditional Heteroskedastic (ARCH) model. In the ARCH model, the conditional variance $Q_t$ was dependent on the squared value of the past residuals from the equations 1:

$$Qt = Vc + \sum_{j=1}^{h} Vj\, e_{t-j}^2 \tag{2}$$

Bollerslev (1987) further developed the framework discussed in ARCH by constructing variance as a function of the lagged values of Qt and the lagged values of $e_t^2$ from the return equation. These types of models became known as the Generalized ARCH (GARCH) models. Based on equation 2, GARCH model can therefore be written as shown below in equation 3:

$$Q_t = Vc + \sum_{j=1}^{h} Vmj Q_{t\text{-}j} + \sum_{j=1}^{r} Vnj\, e_{t-j}^2 \tag{3}$$

In the equation 3, Vc, Vm, and Vn are positive and hence satisfy the criterion for conditional variances (Bollerslev, Chou, & Kroner, 1992). The conditional variances measure the volatility of the model. We can then run the GARCH model by adding the time varying terms to equations 1 as follows:

$$R_{NBI} = \beta_0 + \beta_1(Tuesday) + \beta_2(Wednesday) + \beta_3(Thursday) + \beta_4(Friday) + f\,R_f + \beta_6(R_m\text{-}R_f) + sSMB + hHML + uUMD + \sum_{l=1}^{n} biR_{t-1} + \gamma Q_{t\text{-}l} + et \tag{4}$$

*Where,*

$$Q^2{}_t = Vc + V_1(Tuesday) + V_2(Wednesday) + V_3(Thursday) + V_4(Friday) + V5\,R_f + V_6(R_m\text{-}R_f) + V_7SMB + V_8HML + V_9UMD + V_a Q_{t-1}^2 + V_b e_{t-1}^2 \tag{5}$$

In the equation above, the coefficients for $V_a$ *and* $V_c$ should always be positive, but Vb can be either a positive or a negative fraction. Another feature of financial asset returns is that the increase in volatility in these asset classes is greater in the events of losses than it is in the events of gains. This is known as asymmetry (Nelson, 1991; Wang & Wang, 2011).

According to Black (1976), a drop in the price of an asset reduced its equity, and increased its the debt to equity ratio. This is known as the leverage effect. This made the asset riskier and contributed to the increase in its volatility. Conversely, the volatility of an asset reduced in the presence of positive shocks. It is important to note that although all models with leverage are asymmetric, not all models with asymmetry have leverage. Bouchaud (2001) found that the leverage effect was stronger in the case of indices than it was for individual asset classes.

The GARCH (1, 1) process suggested by Bollerslev (1987) assumes that the distributions are Gaussian, and does not account for the presence of asymmetry. To account for the presence of asymmetry and leverage, Engle (2001) suggested the use of asymmetric GARCH models. However, McAleer (2014) demonstrated that it is not possible to show leverage with asymmetric models such as EGARCH and GJR.

The additional analyses in this study include the use these of two asymmetric GARCH models mentioned in the previous paragraph. These models are the Exponential GARCH (EGARCH) model developed by Nelson (1991), and the GJR (Threshold GARCH) model developed by Glosten, Jaganathan, and Runkle (1993). Incorporating the conditional variance at time t the EGARCH equation is shown below :

$$Log\ (Q^2_t){=}Vc{+}V_1(Tuesday){+}V_2(Wednesday){+}V_3(Thursday){+}V_4(Friday){+}V5\ R_f{+}\ V_6(R_m{-}R_f){+}$$
$$V_7SMB{+}V_8HML{+}V_9UMD{+}\ V_a\log(Q^2_{t-1})\ +\ V_b|e_{t-1}/Q_{t-1}|{+}\xi(e_{t-1}/Q_{t-1}) \qquad (6)$$

In equation 6, the term $|e_{t-1}/Q_{t-1}|$ is the conditional variance, and $(e_{t-1}/Q_{t-1})$ is the lagged value of the residual over conditional standard deviation. $Log(Q^2_{t-1})$ is the logged value of conditional variance. Asymmetry is present when $\xi > 0$. However according to McAleer (2014) and McAleer and Hafner (2014), leverage is not possible in the EGARCH model as both variances of the stochastic processes: $V_b$ and $\xi$ must be positive. Similarly, the GJR process is shown in equation 7 below:

$$Q^2_t{=}Vc{+}V_1(Tuesday){+}V_2(Wednesday){+}V_3(Thursday){+}V_4(Friday){+}V5\ R_f{+}\ V_6(R_m{-}R_f){+}$$
$$V_7SMB{+}V_8HML{+}V_9UMD{+}\ V_aQ^2_{t-1}{+}V_b e^2_{t-1}\ +\gamma\ e^2_{t-1}d_{t-1} \qquad (7)$$

In equation 7, $d_{t-1}$ takes the value of 1 if $u_{t-1}$ is negative, and 0 otherwise. The effect is asymmetric when $\gamma > 0$. However, as demonstrated by McAleer (2014), it is not possible to show leverage using the GJR specification as both variances of the stochastic processes: $V_b$ and the $\gamma$ need to be positive.

## 4. Results

### 4.1 Descriptive Statistics

Table 1 shows the descriptive statistics and t-tests for the day-of-the-week returns for the NBI and its comparison to the market index (S&P 500). The results show that between 2002 and 2012, the NBI returns on Mondays were significantly lower than that of the S&P 500 index. This association was also significant over a shorter period from January 1, 2002 to March 30, 2010. However, there was no significant difference between the S&P 500 and NBI returns on Mondays after March 30, 2010. Also, not surprisingly, across all days and periods, the small company stock dominated NBI returns had a higher standard deviation than the S&P 500 index.

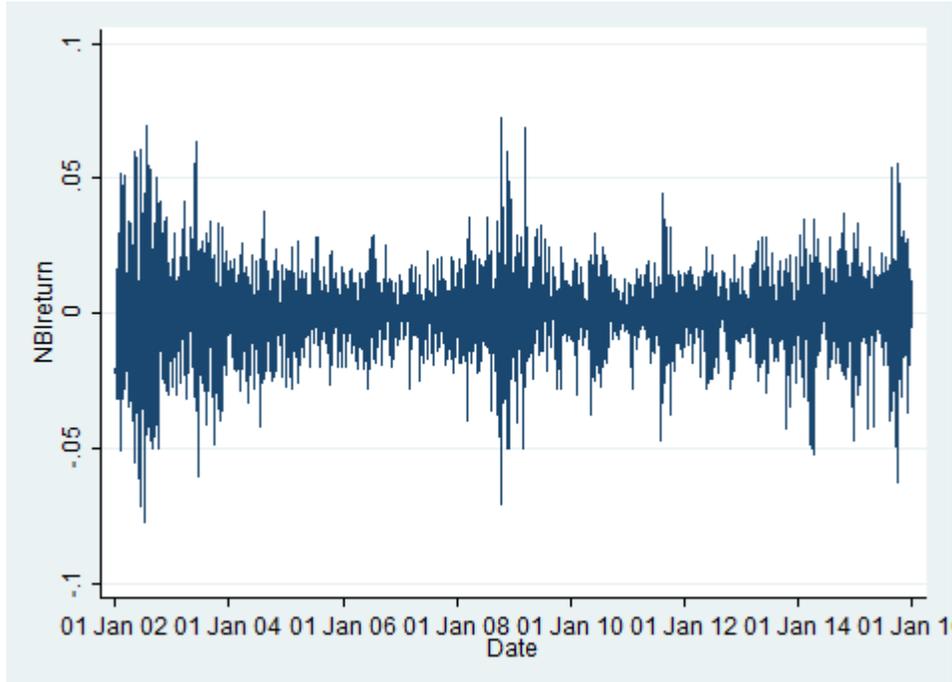

Figure 1: NBI daily returns 2002-2015

The returns from figure 1 show the daily NBI returns during 2002–2015. The volatility in NBI returns can be observed from this graph in 2002, 2008, and 2015. These coincide with periods of volatility for the market in general as shown in figure 2 below:

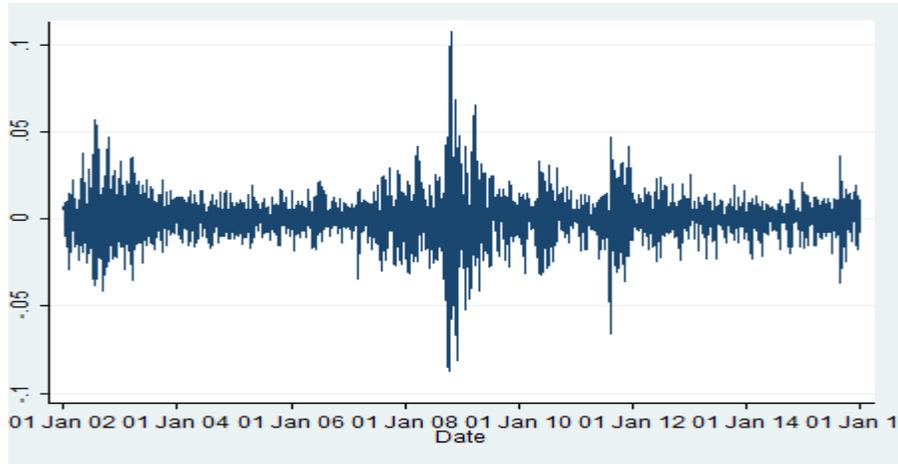

Figure 2: S&P 500 daily returns 2002-2015

Figure 3 below shows the closing prices for the market (S&P 500) and the NBI indices from 2002 through the end of 2015. The figure shows the price increase and the changes in price patterns for the two indices over this period.

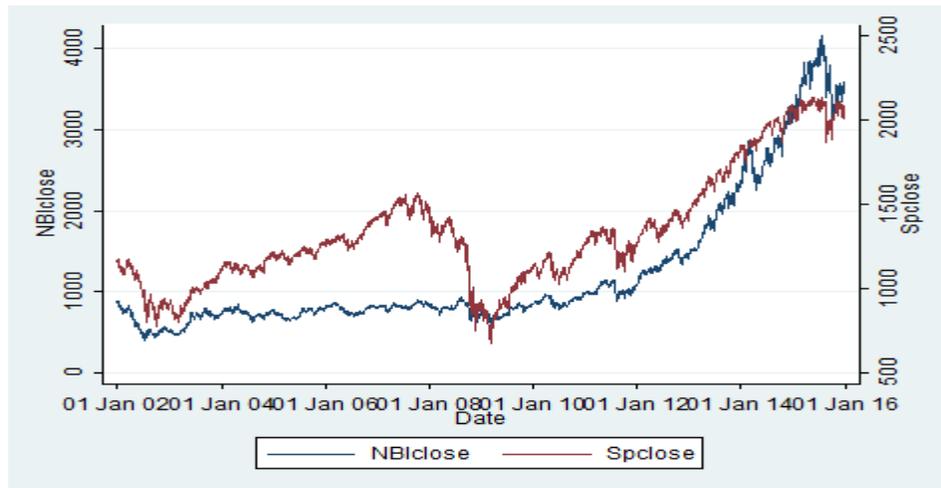

Figure 3: Comparison of S&P 500 and NBI Daily closing prices 2002–2015

### 4.2 Day-of-the-week effect 2002-2015

Results for the day-of-the-week effect are presented in table 2. The first three columns of table 2 include the results from the GARCH (1,1) process. Similarly, the next three columns include results from the EGARCH (1,1) and the last three columns from table 2 include results from the computed GJR model. The results indicate that when compared with Monday, the returns for the NBI were higher on Wednesday, Thursday, and Friday across all three types of GARCH processes over the entire period of this study. Interestingly, the SMB and Rm-Rf variables were positively associated with the biotechnology index returns across all three types of GARCH models. Additionally the momentum (UMD) factor was positively associated and the HML variable was negatively associated with the biotechnology index returns. The lagged returns were significant across all three models.

### 4.3 Day-of-the-week effect during the Pre-ACA Period

Results for the day-of-the-week effect for the pre_ACA period ending March, 30[th], 2010 are presented in table 3. The first three columns of table 3 include the results from the GARCH (1,1) process. Similarly, the next three columns include results from the EGARCH (1,1) and the last three columns from table 2 include results from the computed GJR model. The results indicate that when compared with Monday, the returns for the NBI were higher on Wednesday, Thursday, and Friday across all three types of GARCH processes over the entire period of this study. The SMB and Rm-Rf variables were aslo positively associated with the biotechnology index returns across all three types of GARCH models. Additionally the HML variable was negatively associated with the biotechnology index returns. The lagged returns were significant across all three models.

### 4.4 Day-of-the-week effect during the Post-ACA Period

Results for the day-of-the-week effect in the post-ACA era are presented in table 4. The first three columns of table 4 include the results from the GARCH (1,1) process. Similarly, the next three columns include results from the EGARCH (1,1) and the last three columns from table 4 include results from the computed GJR model. The results indicate that the SMB and Rm-Rf variables were positively associated with the biotechnology index returns across all three types of GARCH models. Additionally the HML variable was negatively associated with the biotechnology index returns. In the EGARCH model the $\gamma$ was significant and negative.

## 5. Conclusion

The findings from this study indicate the presence of a day-of-the-week effect for the biotechnology index during 2002–2015. Another interesting finding is that the significance of higher returns for Wednesday, Thursday, and Friday over the returns for Mondays persisted for the biotechnology stocks over the 2002–2015 period even after controlling for the risk and momentum factors, as suggested by Fama and French (2004) and Carhart (1997). During the period after March 30, 2010, after ACA was signed and the Federal Reserve's first round of quantitative easing program forced interest rates to historic lows, the day-of-the-week effect disappeared. More research is needed in this area to examine whether this change in the behavior of the biotechnology index was due to the related policies that went into effect or to the change in economic cycle as the economy recovered from the great recession. In the future, studies of market efficiency, perhaps using an event study approach, could better extrapolate the effect of the individual policy implications on the price movements of the biotechnology stocks. It is also possible that the period after March 30, 2010 is not long enough to fully estimate the day-of-the week effect. Also, as a comparison in the GARCH models for the S&P 500 index returns that was analyzed for this period[1], no significant day-of-the-week effect were observed after March 30, 2010. This indicates that the macroeconomic changes associated with this period perhaps played a role in the elimination of day-of-the-week effect from the daily returns in the market after March 30, 2010. Positive association between SMB and UMD factors with the biotechnology index returns during this period correlate with momentum of the market and a bias toward the performance of small firms that dominate the NASDAQ biotechnology index.

The findings of this study have implications for active portfolio managers and market participants. The study indicates that although portfolio managers may consider the unique price movements of biotechnology stocks, which usually fall lower than the market on Mondays, and relative to Monday, generate higher returns on Wednesdays, Thursdays, and Fridays in the long run, there does not appear to be any significant day-of-the-week effect present in biotechnology stocks after March 30, 2010.

The positive association of the small firm effect (SMB) and momentum (UMD) factors with the daily biotechnology stock returns indicate that short-term traders may have been able to take advantage of market momentum and small firm price movements both in the long term and the current period. Given the volatility and high reliance on pipeline product development and patent expiration on price valuation of biotechnology stocks, it is recommended that small investors either seek professional investment advice or pursue a buy-and-hold strategy of a diversified biotechnology index fund within their portfolio to benefit from long-term growth expectations.

The findings of this study also have implications for market efficiency which would assume the expected return for the biotechnology stocks would be same each day of the week after controlling for the risks, firm size, price movements of the market, and the general momentum of stocks. However, the day -of-the-week effect persisted for the biotechnology stocks during 2002–2015 even after controlling for the risk- and momentum-related factors, suggesting unstable market dynamics and opportunities for active managers (Lo, 2004).

---

[1] For brevity the GARCH models testing the day-of-the-week effect on S&P 500 index daily returns for the period after March 30, 2010 are not reported separately.

The higher cost of the newly approved drugs launched by the biotechnology companies has become an important consumer issue and has recently been a subject of discussion in the US Presidential debates with some candidates threatening legislation to curb the costs of some of the drugs. Consequently, there is a growing risk of legislation that may result in lower profit margins for biotech firms. In the light of this and other recent policy changes, this study needs to be replicated again in a few years and over more stable economic periods to examine the similarities and dissimilarities in the performance of biotechnology stocks observed in this study.

Table 1: Descriptive Statistics and t-tests

| Day of the week | Period 2002-2015 | | | Jan 01, 2002- March 30, 2010 | | | April 01, 2010- Dec 31, 2015 | | |
|---|---|---|---|---|---|---|---|---|---|
| | NBI | S&P 500 | Difference | NBI | S&P 500 | Difference | NBI | S&P 500 | Difference |
| Monday | | | | | | | | | |
|   Mean(%) | -0.160% | -0.030% | -0.130% | -0.200% | -0.010% | -0.190% | -0.070% | -0.020% | -0.050% |
|   St. Dev | 0.060% | 0.050% | 0.050% | 0.080% | 0.080% | 0.060% | 0.080% | 0.060% | 0.080% |
|   T-test | -2.679 | | | -3.016 | | | -0.5415 | | |
|   Significance | *** | | | *** | | | | | |
|   N | 663 | 663 | 663 | 390 | 390 | 390 | 273 | 273 | 273 |
| Tuesday | | | | | | | | | |
|   Mean(%) | 0.040% | 0.060% | -0.020% | 0.000% | 0.040% | -0.040% | 0.090% | 0.080% | 0.010% |
|   St. Dev | 0.050% | 0.050% | 0.050% | 0.070% | 0.060% | 0.060% | 0.070% | 0.050% | 0.060% |
|   T-test | -0.425 | | | -0.668 | | | 0.2340 | | |
|   Significance | | | | | | | | | |
|   N | 723 | 723 | 723 | 426 | 426 | 426 | 297 | 297 | 297 |
| Wednesday | | | | | | | | | |
|   Mean(%) | 0.050% | 0.040% | 0.010% | 0.080% | 0.050% | 0.030% | 0.010% | 0.030% | -0.020% |
|   St. Dev | 0.050% | 0.040% | 0.040% | 0.080% | 0.060% | 0.060% | 0.070% | 0.060% | 0.060% |
|   T-test | 0.269 | | | 0.549 | | | -0.302 | | |
|   Significance | | | | | | | | | |
|   N | 725 | 725 | 725 | 426 | 426 | 426 | 299 | 299 | 299 |
| Thursday | | | | | | | | | |
|   Mean(%) | 0.010% | 0.030% | -0.020% | 0.020% | 0.010% | 0.010% | 0.010% | 0.080% | -0.070% |
|   St. Dev | 0.060% | 0.050% | 0.040% | 0.080% | 0.060% | 0.060% | 0.080% | 0.060% | 0.060% |
|   T-test | -0.514 | | | 0.207 | | | -1.1820 | | |
|   Significance | | | | | | | | | |
|   N | 709 | 709 | 709 | 418 | 418 | 418 | 291 | 291 | 291 |
| Friday | | | | | | | | | |
|   Mean(%) | -0.010% | 0.030% | -0.010% | -0.070% | 0.000% | -0.070% | 0.080% | 0.010% | 0.070% |
|   St. Dev | 0.050% | 0.030% | 0.040% | 0.070% | 0.050% | 0.060% | 0.070% | 0.050% | 0.060% |
|   T-test | -0.3686 | | | -1.235 | | | 0.9790 | | |
|   Significance | | | | | | | | | |
|   N | 705 | 705 | 705 | 415 | 415 | 415 | 290 | | |

*** *p<.001, **p<0.01, *p<0.05*

Table 2: *GARCH estimates for for day-of-the-week effect 2002-2015*

| NBIreturn | GARCH | | | EGARCH | | | GJR | | |
|---|---|---|---|---|---|---|---|---|---|
| | Coef. | Std. Err. | Sig | Coef. | Std. Err. | Sig | Coef. | Std. Err. | Sig |
| Tue | 0.0006 | 0.0004 | | 0.0006 | 0.0004 | | 0.0005 | 0.0004 | |
| Wed | 0.0009 | 0.0004 | * | 0.0011 | 0.0004 | * | 0.0008 | 0.0003 | * |
| Thu | 0.0010 | 0.0004 | * | 0.0009 | 0.0004 | * | 0.0009 | 0.0003 | * |
| Fri | 0.0009 | 0.0004 | * | 0.0009 | 0.0004 | * | 0.0009 | 0.0004 | * |
| Rf | -2.2332 | 1.7477 | | -3.6231 | 2.5226 | | -3.232 | 2.6771 | |
| Rm-Rf | 0.7085 | 0.0117 | *** | 0.7252 | 0.0123 | *** | 0.7071 | 0.0116 | *** |
| SMB | 0.4594 | 0.0261 | *** | 0.4477 | 0.0253 | *** | 0.4608 | 0.0261 | *** |
| HML | -0.6225 | 0.0257 | *** | -0.6344 | 0.0266 | *** | -0.6242 | 0.0255 | *** |
| UMD | 0.0738 | 0.0169 | *** | 0.0748 | 0.0163 | *** | 0.0726 | 0.0169 | *** |
| $R_{NBI-1}$ | 0.0481 | 01279 | ** | 0.0482 | 0.0113 | *** | 0.0479 | 0.0112 | *** |
| *Variance Equation* | | | | | | | | | |
| a | 0.0841 | 0.0081 | *** | 0.0676 | 0.0131 | *** | 0.0886 | 0.0102 | *** |
| b | 0.9082 | 0.0083 | *** | 0.9051 | 0.0091 | *** | 0.8981 | 0.0082 | *** |
| γ | | | | -0.0095 | 0.0080 | | 0.0076 | 0.0118 | |
| Intercept | 0.0001 | 0.0000 | *** | -0.0306 | 0.0082 | *** | 0.0000 | 0.0000 | *** |

*** $p<.001$, ** $p<0.01$, * $p<0.05$

Table 3: *GARCH estimates for for day-of-the-week effect Pre-ACA*

| NBIreturn | GARCH | | | EGARCH | | | GJR | | |
|---|---|---|---|---|---|---|---|---|---|
| | Coef. | Std. Err. | Sig | Coef. | Std. Err. | Sig | Coef. | Std. Err. | Sig |
| Tue | 0.0006 | 0.0006 | | 0.0005 | 0.0005 | | 0.0005 | 0.0004 | |
| Wed | 0.0013 | 0.0005 | * | 0.0012 | 0.0005 | * | 0.0008 | 0.0003 | * |
| Thu | 0.0012 | 0.0005 | * | 0.0012 | 0.0005 | * | 0.0008 | 0.0003 | * |
| Fri | 0.0011 | 0.0004 | * | 0.0010 | 0.0004 | * | 0.0007 | 0.0003 | * |
| Rf | -2.6367 | 2.3226 | | -3.4877 | 2.8982 | | -2.9781 | 1.754 | |
| Rm-Rf | 0.7675 | 0.0163 | *** | 0.7758 | 0.0163 | *** | 0.7241 | 0.0125 | *** |
| SMB | 0.5304 | 0.0313 | *** | 0.5214 | 0.0324 | *** | 0.4515 | 0.0257 | *** |
| HML | -0.5605 | 0.0342 | *** | -0.5623 | 0.0341 | *** | -0.6219 | 0.0214 | *** |
| UMD | 0.0339 | 0.0209 | | 0.0367 | 0.0200 | | 0.0631 | 0.0712 | |
| $R_{NBI-1}$ | 0.0731 | 0.0241 | ** | 0.0656 | 0.0136 | *** | 0.0484 | 0.0117 | *** |
| *Variance Equation* | | | | | | | | | |
| a | 0.0700 | 0.009 | *** | 0.0312 | 0.0093 | *** | 0.0679 | 0.0009 | *** |
| b | 0.9251 | 0.009 | *** | 0.9327 | 0.0026 | *** | 0.9009 | 0.0082 | *** |
| $\gamma$ | | | | -0.0088 | 0.0112 | | 0.0076 | 0.0126 | |
| Intercept | 0.0001 | 0.0000 | *** | -0.0616 | 0.0033 | *** | 0.0000 | 0.0000 | *** |

*** p<.001, **p<0.01, *p<0.05

Table 4: *GARCH estimates for for day-of-the-week effect Post-ACA*

| NBIreturn | GARCH | | | EGARCH | | | GJR | | |
|---|---|---|---|---|---|---|---|---|---|
| | Coef. | Std. Err. | Sig | Coef. | Std. Err. | Sig | Coef. | Std. Err. | Sig |
| Tue | 0.0009 | 0.0007 | | 0.0010 | 0.0008 | | 0.0010 | 0.0007 | |
| Wed | 0.0005 | 0.0007 | | 0.0012 | 0.0008 | | 0.0007 | 0.0006 | |
| Thu | 0.0010 | 0.0007 | | 0.0012 | 0.0005 | | 0.0011 | 0.0008 | |
| Fri | 0.0008 | 0.0007 | | 0.0010 | 0.0007 | | 0.0007 | 0.0005 | |
| Rf | -6.580 | 9.9071 | | -5.6771 | 5.0241 | | -8.1116 | 5.625 | |
| Rm-Rf | 0.5731 | 0.0192 | *** | 0.5473 | 0.0172 | *** | 0.5694 | 0.0174 | *** |
| SMB | 0.4141 | 0.0453 | *** | 0.4395 | 0.0408 | *** | 0.4035 | 0.0445 | *** |
| HML | -0.7035 | 0.0474 | *** | -0.6583 | 0.0341 | *** | -0.6844 | 0.0453 | *** |
| UMD | 0.2841 | 0.0332 | *** | 0.3001 | 0.0317 | *** | 0.2984 | 0.0332 | *** |
| $R_{NBI-1}$ | 0.0162 | 0.0200 | | 0.0258 | 0.0193 | | 0.0197 | 0.0242 | |
| *Variance Equation* | | | | | | | | | |
| a | 0.0912 | 0.0162 | *** | 0.0512 | 0.0093 | *** | 0.0729 | 0.0212 | *** |
| b | 0.8904 | 0.0132 | *** | 0.9368 | 0.0008 | *** | 0.8842 | 0.0159 | *** |
| $\gamma$ | | | | -0.0106 | 0.0037 | ** | 0.0248 | 0.0256 | |
| Intercept | 0.0000 | 0.0000 | ** | -0.0518 | 0.0082 | *** | 0.0000 | 0.0000 | *** |

*** $p<.001$, ** $p<0.01$, * $p<0.05$